%% file: main.tex
\documentclass[a4paper,11pt]{article}
\usepackage{subcaption}
\usepackage{pos}

\title{Status and performance of mass-produced IceCube Upgrade mDOMs}
\ShortTitle{Mass-produced IceCube Upgrade mDOMs}

\emailAdd{satoshi.fukami@desy.de}

\author{The IceCube Collaboration \\{\normalsize \normalfont(a complete list of authors can be found at the end of the proceedings)}\\}

\abstract{
IceCube Upgrade, starting in the Antarctic summer season 2025/2026, will enhance the sensitivity of the current IceCube in the GeV range and improve understanding of the ice properties. Around 700 new modules will be deployed deep in the ice along seven strings, spaced $3\,\mathrm{m}$ vertically within a horizontal region of $100\,\mathrm{m}$. IceCube Upgrade mainly consists of two types of Cherenkov photon detectors, $\sim280$ Dual optical sensors in an Ellipsoid Glass for Gen2 (D-Eggs) and $\sim400$ multi-PMT Digital Optical Modules (mDOMs).
Here, we present the status of mDOM production and acceptance testing for mass-produced mDOMs. The majority of the modules have already been produced, and all 128 modules on the first 2 strings have been successfully shipped to the South Pole after multiple steps of testing. Each mDOM contains 24 photomultipliers nearly isotropically distributed on the surface to achieve uniform photon sensitivity, making the integration and testing complicated. To reduce the amount of testing time, we have developed parallelized and automated software.

\vspace{4mm}

{\bfseries Corresponding authors:}
Satoshi Fukami$^{1*}$\\
{$^{1}$ \itshape Deutsches Elektronen-Synchrotron}\\[4mm]
$^*$ Presenter
}

\ConferenceLogo{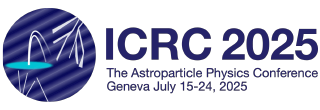}

\FullConference{39th International Cosmic Ray Conference (ICRC2025)\\
 15–24 July 2025\\
Geneva, Switzerland\\}

\begin{document}

\maketitle

\section{Introduction}
IceCube Upgrade is the expansion of the current IceCube Neutrino Observatory to detect relatively low-energy neutrinos down to GeV. This extension is expected to significantly advance multiple science cases such as the neutrino oscillation study and the astrophysical transients \cite{Upgrade_ICRC2019,Upgrade_transient_science}.
To detect dim and compact Cherenkov light yield originating from GeV neutrinos, optical modules will be deployed more densely in both vertical and horizontal directions. 
Optical modules and other calibration modules will be distributed along 7 strings with a typical vertical spacing of $\sim$$3\,\mathrm{m}$ within a horizontal diameter of $\sim$$100\,\mathrm{m}$. These values are much smaller than the existing IceCube geometry. Due to its density, IceCube Upgrade will allow further investigation of the ice properties, which is the major contributor to the uncertainty of the scientific results of IceCube.

\begin{figure}[ht]
    \centering
    \begin{subfigure}{.45\textwidth}
      \centering
      \includegraphics[width=.75\linewidth]{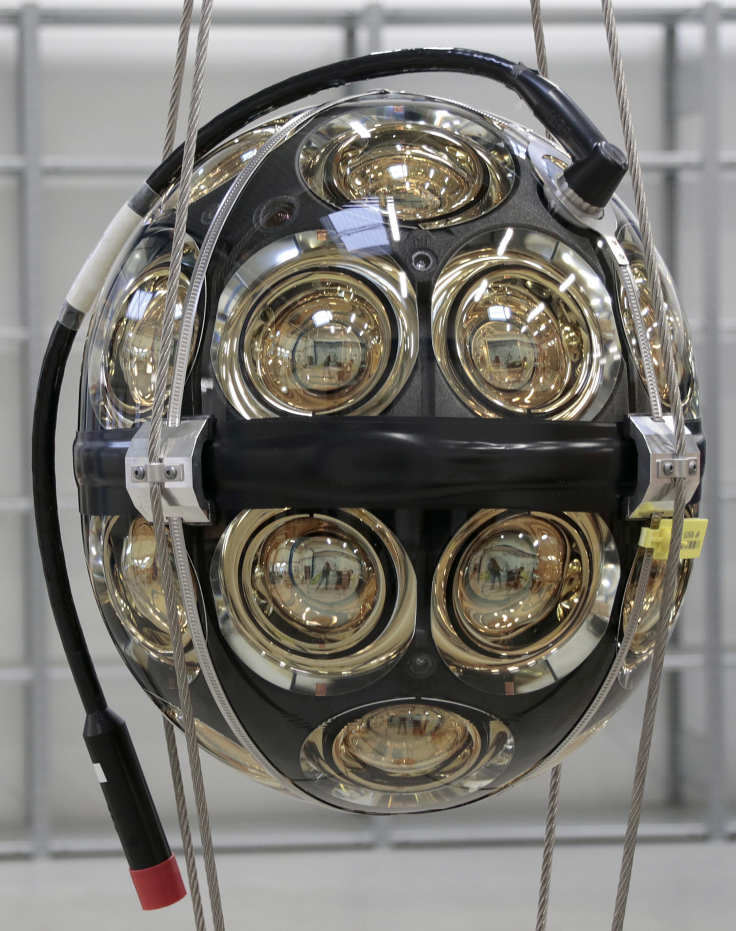}
      \label{fig:mDOM_image}
    \end{subfigure}
    \begin{subfigure}{.45\textwidth}
      \centering
      \includegraphics[width=.9\linewidth]{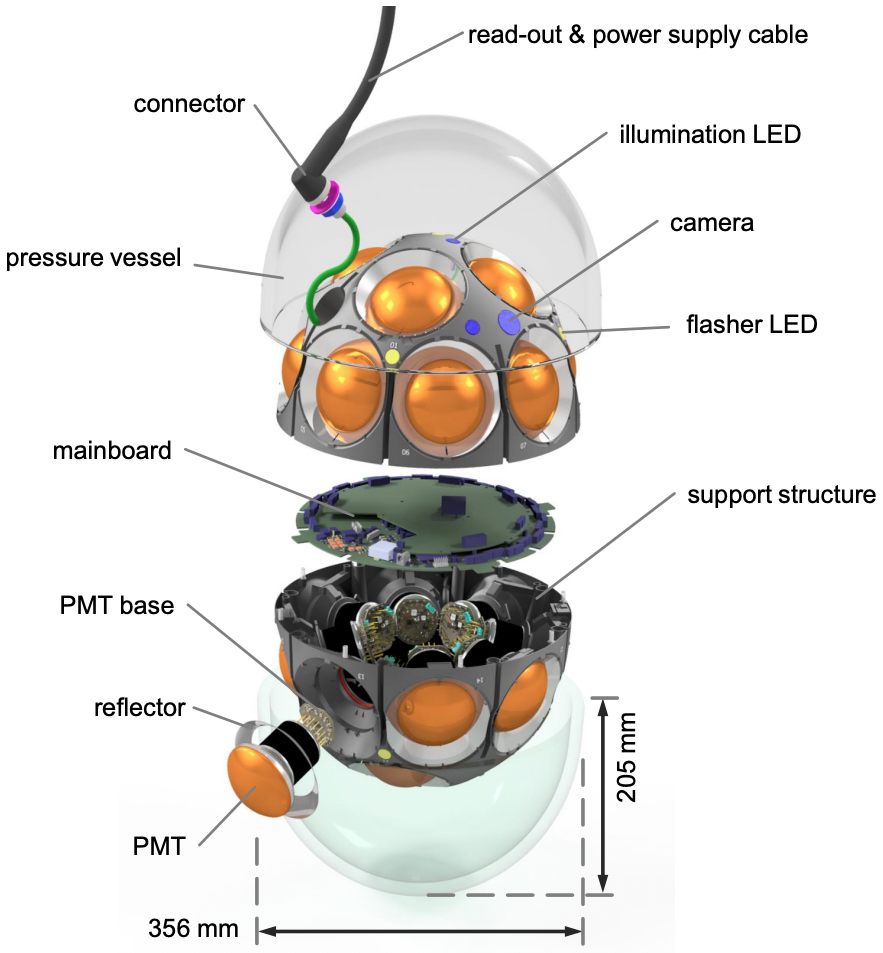}
      \label{fig:mDOM_expanded}
    \end{subfigure}
    \caption{Left: mDOM harnessed by steel ropes (credit: S.Niedworok/DESY). Right: Expanded view of an mDOM \cite{mDOM_ICRC2021}.}
    \label{fig:mDOM}
\end{figure}

Optical modules are also updated for IceCube Upgrade to have better performance. There are two major types of optical modules for IceCube Upgrade: $\sim$280 "Dual optical sensors in an Ellipsoid Glass for Gen2" (D-Eggs) \cite{Degg_paper} and $\sim$400 "multi-PMT Digital Optical Modules" (mDOMs) \cite{mDOM_ICRC2021}. Each type of module has multiple photomultiplier tubes (PMTs) with better quantum efficiencies than IceCube modules to increase the photon sensitivity averaged over the directions \cite{Gen2_modules_ICRC2017}. The mDOM consists of 24 3-inch PMTs to achieve nearly isotropic photon acceptance (see Fig.~\ref{fig:mDOM}). It has a low-power-consumption electronics to stream 24 PMT signals without dead time. A set of environmental sensors are embedded in the electronics board. Additionally it contains auxiliary devices such as wide-field cameras and flasher LEDs used for PMT calibration and ice monitoring.

The module deployment of all the 7 Upgrade strings is planned in the austral summer of 2025/2026. mDOMs have been intensely manufactured, tested, and evaluated towards the deployment. In this study, we present the status of production and testing of the mDOMs to be shipped and deployed at the South Pole.

\section{Status of Production and Testing}

The mDOMs have been produced at two sites: Deutsches-Electronen Synchrotron (DESY) in Zeuthen and Michigan State University (MSU). Components such as PMTs, glass vessels, electronics boards, and calibration devices, are transported from respective initial testing sites or production sites to the two assembly sites above. DESY and MSU are supposed to produce $\sim$230, $\sim$205 mDOMs including spares respectively. The production of all the modules has been done as of June 2025. 

After the assembly of the modules, they go through a suite of in-depth acceptance tests (Final Acceptance Testing, FAT) to confirm the full functionalities of the whole module. The testing procedure is almost the same as described in \cite{mDOM_testing_ICRC2023}. There are two testing sites: the Dark Freezer Lab (DFL) and the optical setup. In DFL FAT, we perform tests on the PMTs and other components such as calibration devices while the temperature is cycled twice between $+20^{\circ}\mathrm{C}$ and $-40^{\circ}\mathrm{C}$. The key tests of the DFL FAT are listed in Table \ref{tab:test_items}. The total duration is roughly 2 weeks with the current configuration, dominated by the long-term dark rate monitoring at $-40^{\circ}\mathrm{C}$ for 5 days and the waiting time for temperature stabilization. 

To perform DFL FAT efficiently, we have developed a control software so that many modules, at least 24, can be tested in parallel. Now we are simultaneously running not only as many threads as the number of connected mDOMs but also monitoring threads for temperature, cable connection, and memory usage without any racing conditions. We also managed to control the temperature setting remotely for fully automatized testing. The same software is used for the optical FAT.

In the optical FAT, we mainly measure the linearity and transit times of each PMT channel by precisely controlling the intensity of the light pulse using a UV laser. We can perform only one module at a time due to the setup. It takes in total ~2 hours including redundant measurements such as the gain calibration. The test items are listed in Table \ref{tab:test_items}.

We have finished the FAT for most of the produced modules in both DESY and MSU as of June 2025.

\begin{table}[h!]
\begin{center}
\begin{tabular}{ |l|l| } 
\hline
\multicolumn{1}{|c|}{DFL FAT ($\sim$2 weeks for >=24 modules)} & \multicolumn{1}{|c|}{optical FAT ($\sim$2h for 1 module)}\\
\hline
- PMT gain calibration/monitoring & - PMT gain calibration \\ 
- discriminator threshold calibration/monitoring & - linearity measurement \\
- dark rate monitoring & - transit time measurement \\
- waveform acquisition & - calibration device tests \\
- calibration device tests & - cable swapping check \\ 
- mainboard sensor monitoring & \\
- Simple Test Framework tests (basic functionality check) & \\
\hline
\end{tabular}
\caption{List of main tests for DFL FAT and optical FAT.}
\label{tab:test_items}
\end{center}
\end{table}

\section{Evaluation Procedure}

\begin{figure}[ht]
    \centering
    \begin{subfigure}{.45\textwidth}
      \centering
      \includegraphics[width=\linewidth]{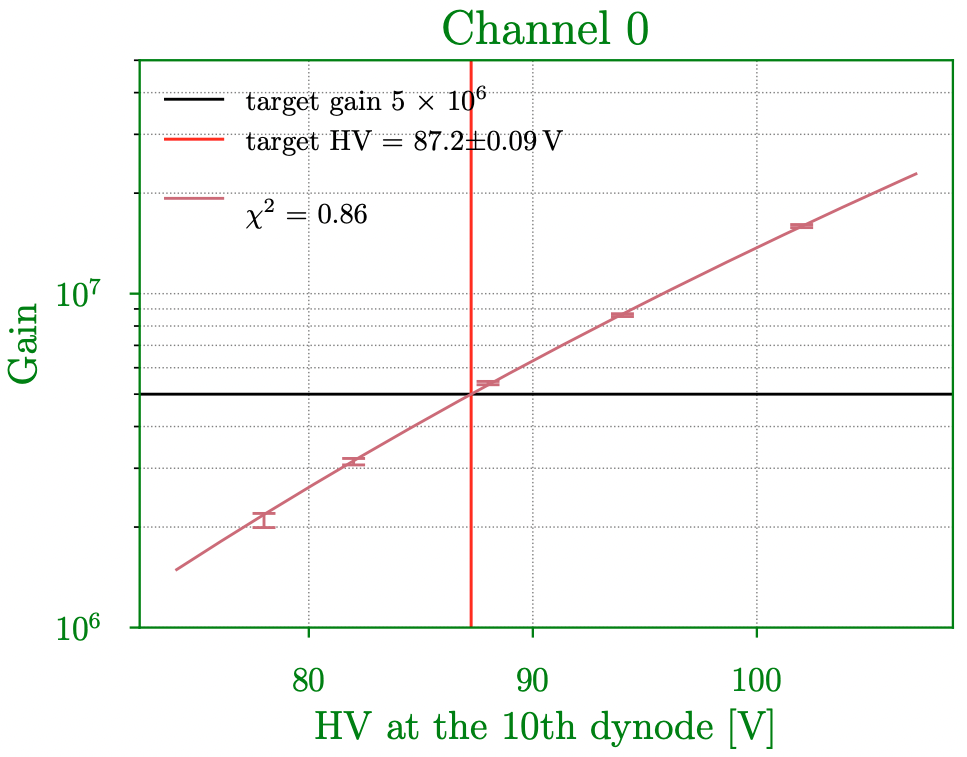}
      \label{fig:summary_pdf}
    \end{subfigure}
    \begin{subfigure}{.5\textwidth}
      \centering
      \includegraphics[width=\linewidth]{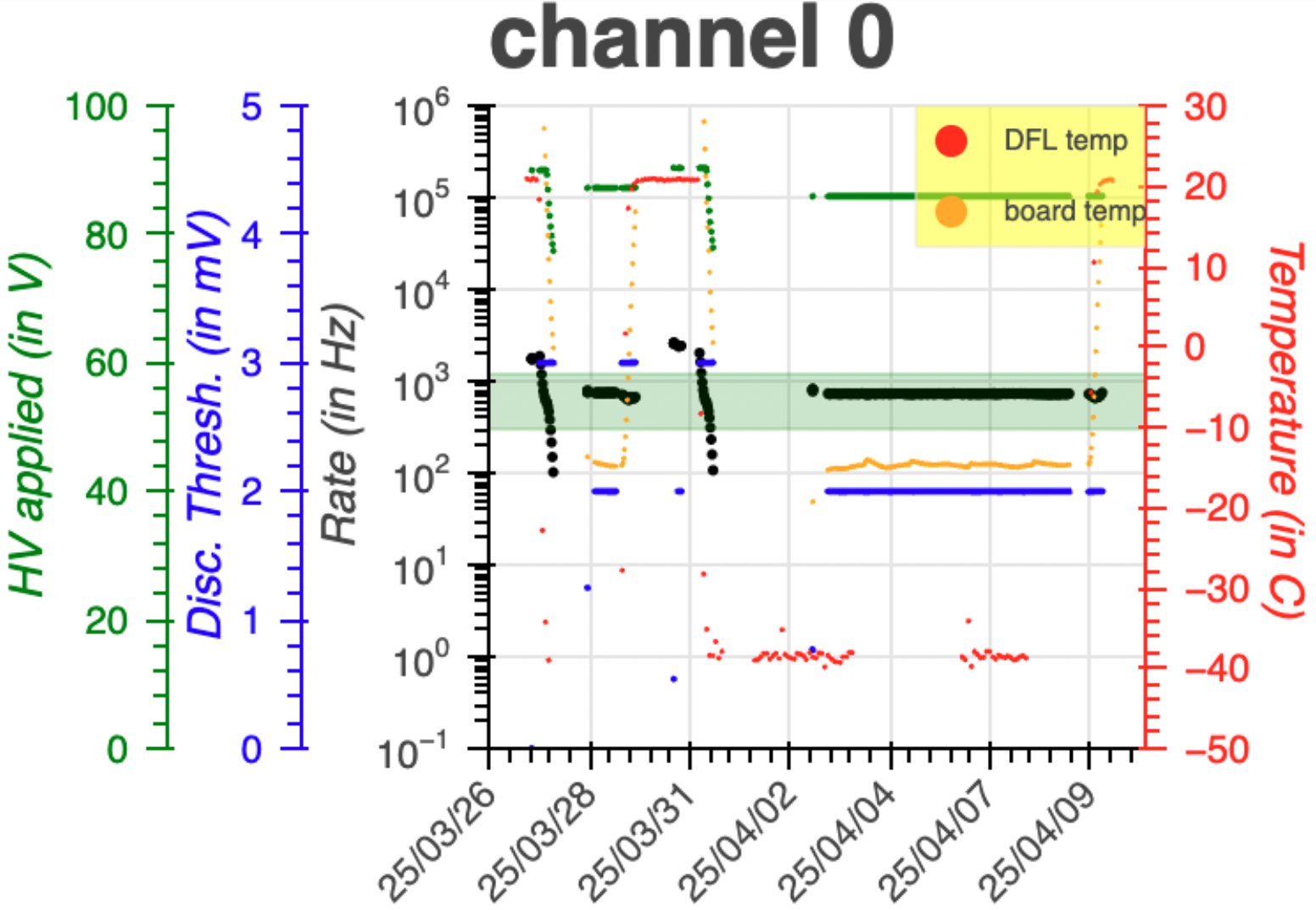}
      \label{fig:summary_html}
    \end{subfigure}
    \caption{Example plots from the two summary files used for the module evaluation. Left: Gain-vs-target voltage curve for an arbitrary PMT channel in the summary pdf file. Right: Dark rate monitoring plot for an arbitrary PMT channel in the summary html file. The temperature, the applied PMT voltage, and the trigger threshold are overplotted to show the configuration stability.}
    \label{fig:summary_pdf_html}
\end{figure}

After the testing, we evaluated the modules. We have produced scripts to compile the results of various test measurements for each mDOM into two summary files: one PDF and one HTML. Fig.~\ref{fig:summary_pdf_html} shows an example plot in each file. The PDF file is the main reference for the evaluation as it includes complete information on all the measurements. The HTML file is produced by the Bokeh Python library and stored in a web server to provide a browser-based interface to the test results. It offers interactivity with the plots stored in separate json files. It is useful for debugging by scaling the plots or highlighting specific measurements. During our evaluation procedure, we relied on these two summary files which are supplemental to each other.

We have adopted strict criteria for modules shipped to the South Pole. For every module, we allow up to only 1 malfunctioning PMT channel. We always require stable PMT dark rates after the temperature is settled down and reliable gain calibration/monitoring at the cold temperature. We also inspected calibration devices and onboard sensors visually using the summary files.

We have required at least two experts to review the same mDOM. For modules with unexpected behaviors of PMTs or calibration devices, we always discussed them among a broader group within the IceCube Collaboration. In DESY, we have finished evaluation for $\sim$210 mDOMs, out of which 194 have passed our criteria as of June 2025.

\section{Results}

Here, we present statistics on key PMT parameters of the 194 successful modules measured during the acceptance testing.

\subsection{Gain Calibration}

We set the gain of every PMT to $5\times10^6$. This value achieves a good balance between the signal-to-noise ratio and the lifespan. The gain is calibrated by measuring charges of single photo-electron (SPE) events with a range of applied voltages and fitting a model curve to the data. We adjusted the flasher LED intensity to the SPE level and flashed them with 10 kHz to quickly sample SPE events using the flasher-synchronized trigger. 

\begin{figure}[ht]
    \centering
    \begin{subfigure}{.5\textwidth}
      \centering
      \includegraphics[width=\linewidth]{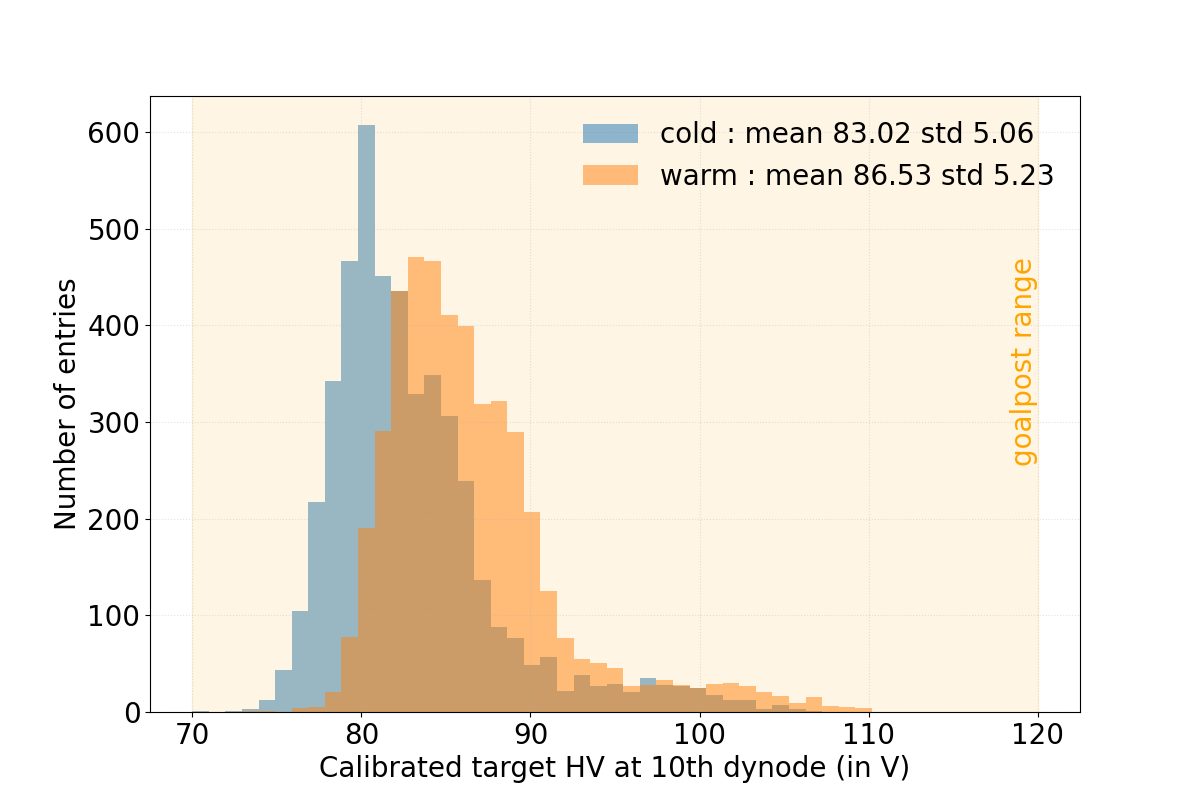}
      \label{fig:targetHV}
    \end{subfigure}
    \begin{subfigure}{.49\textwidth}
      \centering
      \includegraphics[width=\linewidth]{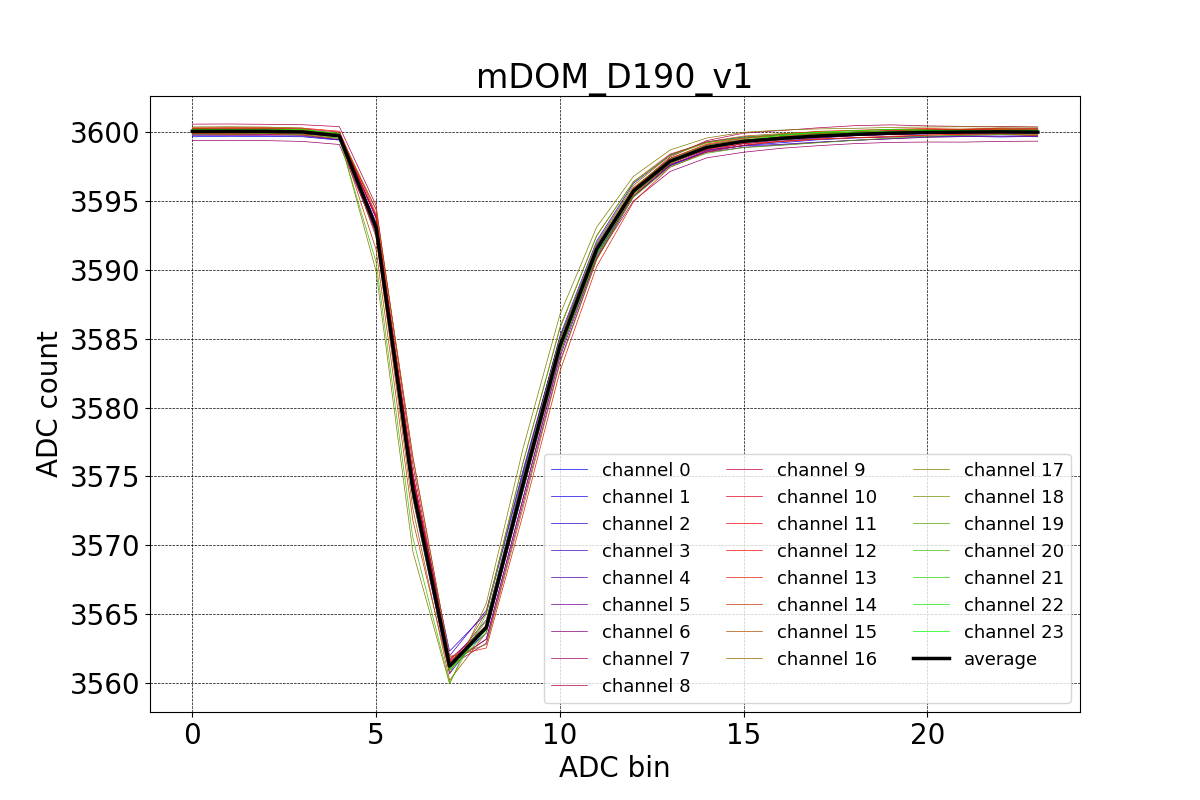}
      \label{fig:SPEwaveform}
    \end{subfigure}
    \caption{Left: Calibrated target voltage distribution applied for PMTs of the 194 mDOMs at the two different temperatures of $+20^{\circ}\mathrm{C}$ and $-40^{\circ}\mathrm{C}$. Right: Averaged SPE waveforms for each PMT channel and its mean of an example mDOM.}
    \label{fig:targetHV_SPEwaveform}
\end{figure}

The left panel of Fig.~\ref{fig:targetHV_SPEwaveform} shows the distributions of the calibrated voltage for all the PMTs of the 194 modules. The voltage shown in the figure is between the anode and the 10th dynode, which is 12 times smaller than the anode-to-cathode voltage. The calibrated voltages are within the expected goalpost range for most of the PMTs. The two distributions are measured at different temperatures of $+20^{\circ}\mathrm{C}$ and $-40^{\circ}\mathrm{C}$. We can see a clear offset originating from the temperature dependency of the PMT dynodes and photocathode. The distributions are also compatible with a previous measurement on bare PMTs done by \cite{mDOM_PMT_paper}.

There are $\sim$10 malfunctioning PMTs which are not shown in this figure. This rate is small compared to the total PMT count of 4500. No modules have multiple dead PMT channels.

The right panel of Fig.~\ref{fig:targetHV_SPEwaveform} shows the averaged digitized waveforms of SPE event signals for an example module. During the gain measurement, charges are automatically calculated by integrating digitized waveforms inside the microcontroller on the electronics board. Although the functionalities of PMT channels can be confirmed by the calculated charges alone, we choose random modules and inspect their waveforms as a crosscheck.

\subsection{Linearity and Transit Time}

\begin{figure}[ht]
    \centering
    \begin{subfigure}{.4\textwidth}
      \centering
      \includegraphics[width=\linewidth]{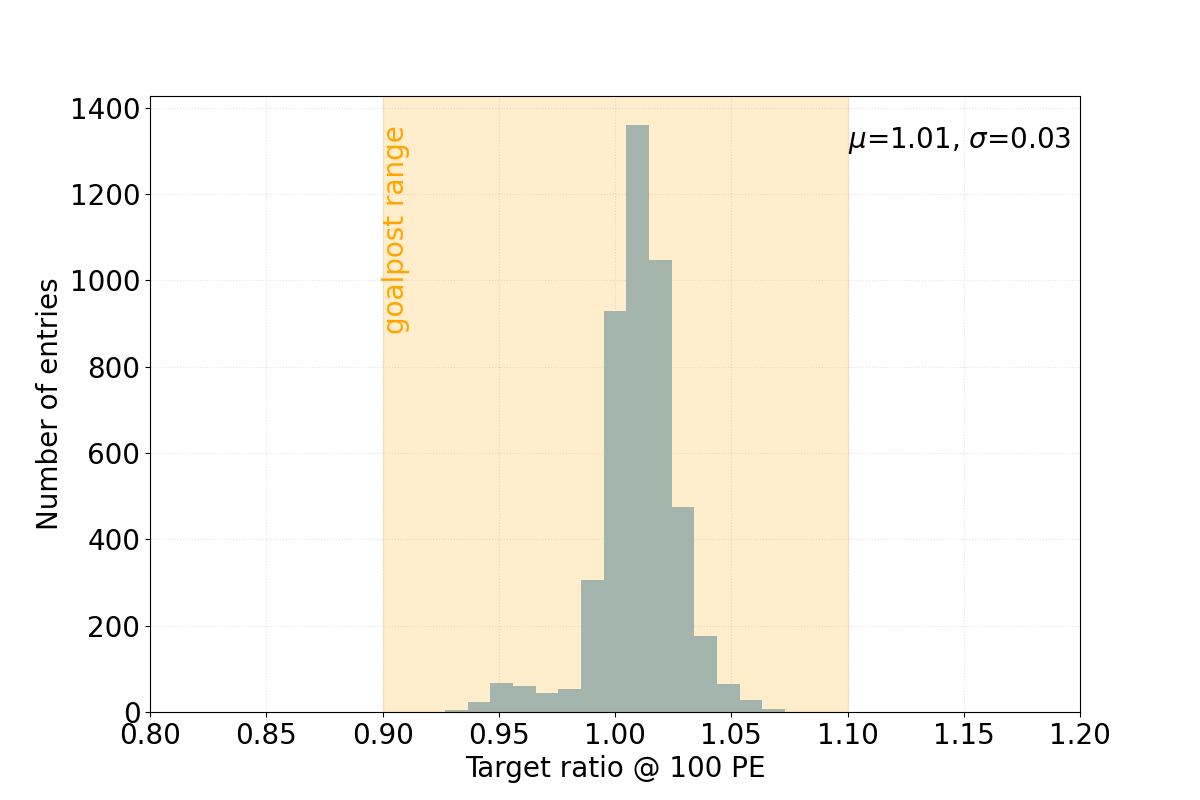}
      \label{fig:linearity}
    \end{subfigure}
    \begin{subfigure}{.58\textwidth}
      \centering
      \includegraphics[width=\linewidth]{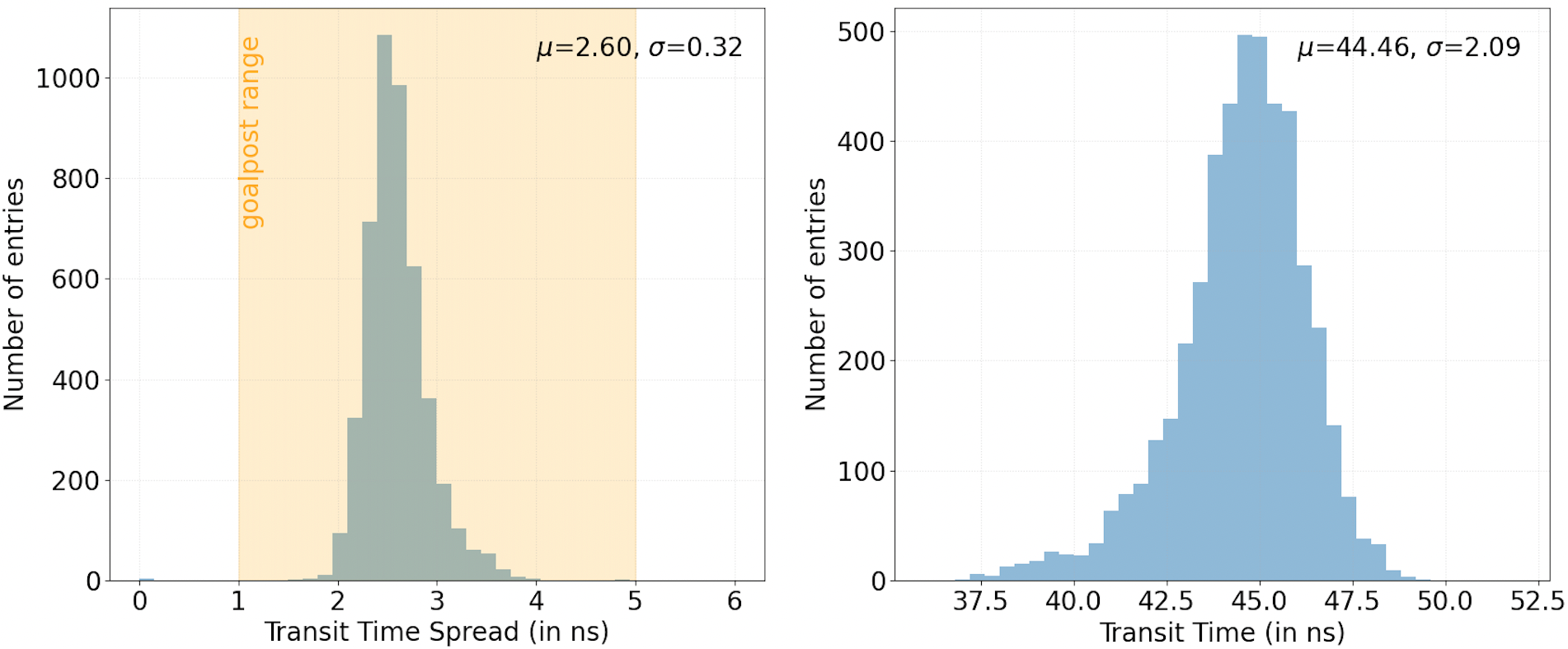}
      \label{fig:TT}
    \end{subfigure}
    \caption{Left: Linearity ratio distribution at 100 PE for the PMTs of the 194 mDOMs. Center, Right: Transit time spread and transit time distributions.}
    \label{fig:linearity_TT}
\end{figure}

The PMT channels of Upgrade optical modules are required to have charge linearity up to 100 PE with the nominal gain. Details of the linearity measurement are described in \cite{mDOM_testing_ICRC2023}. The left panel of Fig.~\ref{fig:linearity_TT} shows the distribution of the ratio between the extrapolation of the linear fitting in the low-charge region and the empirical entire-range fitting at 100 PE. The required range of 10\% is achieved for most of the PMTs. 

Timing resolution is another key parameter of Upgrade PMTs. We measured the transit times of PMTs and their spread in a method described in \cite{mDOM_testing_ICRC2023} with the nominal gain set for each PMT. The center and right panels of Fig.~\ref{fig:linearity_TT} are the distributions of the two parameters. The transit time spread distribution has a peak of $\sim$2.6 ns, well inside the required goalpost region. The transit time distribution is concentrated between 40-50 ns. Although there are no requirements on the transit time itself, the parameters are vital for the timing calibration during in-ice observations.

As in the calibrated voltage distribution, both the linearity and the transit time distributions have $\sim$10 missing entries corresponding to malfunctioning PMTs.

\subsection{Dark Rate Monitoring}

\begin{figure}[ht]
    \centering
    \includegraphics[width=0.6\linewidth]{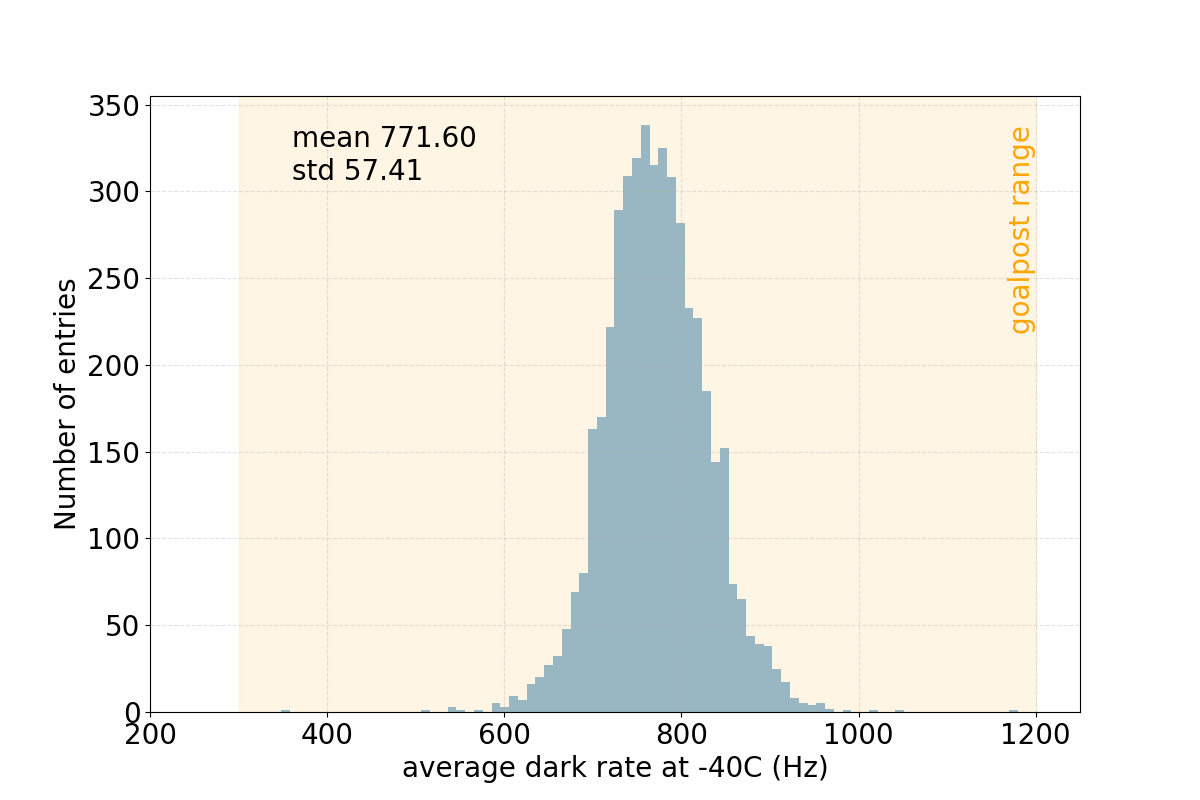}
    \caption{Average dark rate distribution for the PMTs of the 194 mDOMs at the cold temperature of $-40^{\circ}\mathrm{C}$.}
    \label{fig:dark_rate}
\end{figure}

The dark rate was monitored for 5-10 days at $-40^{\circ}\mathrm{C}$ to confirm the stability of the PMT channel performance with the calibrated applied voltage and a fixed discriminator threshold of $\sim$0.4 PE. Fig.~\ref{fig:dark_rate} shows the distribution of the averaged dark rate for each PMT channel. Most of the PMT channels have dark rates of 600-1000 Hz, within the expected range. This average rate of $\sim$800 Hz is significantly larger than the one reported in \cite{mDOM_PMT_paper} due to the additional radioactive decay from the glass vessel besides the PMT itself. The dark rate is also affected by the surrounding medium. In ice, the dark rate is expected to decrease due to the smaller difference in the refractive index from the glass vessel than in the air.

\section{Shipment and Deployment Schedule}

Modules with positive evaluation results are prepared for shipment to the South Pole. They are first harnessed with metal pieces to interface wire ropes used for hanging. After a load-tolerance test, they are packed in carton boxes and shipped. We have already shipped 128 mDOMs from DESY to the South Pole in summer 2024. The rest will be shipped from DESY and MSU in July/August 2025 and will arrive at the South Pole a few months later. From November 2025, mDOMs and all the other Upgrade modules will go through the last acceptance testing at the South Pole before the deployment in the next months.

\section{Summary}

The mDOMs are critical modules for IceCube Upgrade. The production of all the $\sim$430 modules was completed including spare ones, and the acceptance testing has almost finished. The modules have gone through the evaluation process and the results are positive for most of the modules. They will be shipped in the summer of 2025 from DESY and MSU, and will be deployed in the austral summer of 2025/2026 together with the other Upgrade modules.

\clearpage

\input{authorlist_IceCube.tex}

\end{document}

%% file: authorlist_IceCube.tex
\section*{Full Author List: IceCube Collaboration}

\scriptsize
\noindent
R. Abbasi$^{16}$,
M. Ackermann$^{63}$,
J. Adams$^{17}$,
S. K. Agarwalla$^{39,\: {\rm a}}$,
J. A. Aguilar$^{10}$,
M. Ahlers$^{21}$,
J.M. Alameddine$^{22}$,
S. Ali$^{35}$,
N. M. Amin$^{43}$,
K. Andeen$^{41}$
% ...
\\
\\
% ...
$^{10}$ Universit{\'e} Libre de Bruxelles, Science Faculty CP230, B-1050 Brussels, Belgium \\
% ...
$^{16}$ Department of Physics, Loyola University Chicago, Chicago, IL 60660, USA \\
$^{17}$ Dept. of Physics and Astronomy, University of Canterbury, Private Bag 4800, Christchurch, New Zealand \\
% ...
$^{21}$ Niels Bohr Institute, University of Copenhagen, DK-2100 Copenhagen, Denmark \\
$^{22}$ Dept. of Physics, TU Dortmund University, D-44221 Dortmund, Germany \\
% ...
$^{35}$ Dept. of Physics and Astronomy, University of Kansas, Lawrence, KS 66045, USA \\
% ...
$^{39}$ Dept. of Physics and Wisconsin IceCube Particle Astrophysics Center, University of Wisconsin{\textemdash}Madison, Madison, WI 53706, USA \\
% ...
$^{41}$ Department of Physics, Marquette University, Milwaukee, WI 53201, USA \\
% ...
$^{43}$ Bartol Research Institute and Dept. of Physics and Astronomy, University of Delaware, Newark, DE 19716, USA \\
% ...
$^{63}$ Deutsches Elektronen-Synchrotron DESY, Platanenallee 6, D-15738 Zeuthen, Germany \\
$^{\rm a}$ also at Institute of Physics, Sachivalaya Marg, Sainik School Post, Bhubaneswar 751005, India \\
% ...

\subsection*{Acknowledgments}

\noindent
The authors gratefully acknowledge the support from the following agencies and institutions:
USA {\textendash} U.S. National Science Foundation-Office of Polar Programs,
U.S. National Science Foundation-Physics Division,
U.S. National Science Foundation-EPSCoR,
U.S. National Science Foundation-Office of Advanced Cyberinfrastructure,
Wisconsin Alumni Research Foundation,
Center for High Throughput Computing (CHTC) at the University of Wisconsin{\textendash}Madison,
Open Science Grid (OSG),
Partnership to Advance Throughput Computing (PATh),
Advanced Cyberinfrastructure Coordination Ecosystem: Services {\&} Support (ACCESS),
Frontera and Ranch computing project at the Texas Advanced Computing Center,
U.S. Department of Energy-National Energy Research Scientific Computing Center,
Particle astrophysics research computing center at the University of Maryland,
Institute for Cyber-Enabled Research at Michigan State University,
Astroparticle physics computational facility at Marquette University,
NVIDIA Corporation,
and Google Cloud Platform;
Belgium {\textendash} Funds for Scientific Research (FRS-FNRS and FWO),
FWO Odysseus and Big Science programmes,
and Belgian Federal Science Policy Office (Belspo);
Germany {\textendash} Bundesministerium f{\"u}r Forschung, Technologie und Raumfahrt (BMFTR),
Deutsche Forschungsgemeinschaft (DFG),
Helmholtz Alliance for Astroparticle Physics (HAP),
Initiative and Networking Fund of the Helmholtz Association,
Deutsches Elektronen Synchrotron (DESY),
and High Performance Computing cluster of the RWTH Aachen;
Sweden {\textendash} Swedish Research Council,
Swedish Polar Research Secretariat,
Swedish National Infrastructure for Computing (SNIC),
and Knut and Alice Wallenberg Foundation;
European Union {\textendash} EGI Advanced Computing for research;
Australia {\textendash} Australian Research Council;
Canada {\textendash} Natural Sciences and Engineering Research Council of Canada,
Calcul Qu{\'e}bec, Compute Ontario, Canada Foundation for Innovation, WestGrid, and Digital Research Alliance of Canada;
Denmark {\textendash} Villum Fonden, Carlsberg Foundation, and European Commission;
New Zealand {\textendash} Marsden Fund;
Japan {\textendash} Japan Society for Promotion of Science (JSPS)
and Institute for Global Prominent Research (IGPR) of Chiba University;
Korea {\textendash} National Research Foundation of Korea (NRF);
Switzerland {\textendash} Swiss National Science Foundation (SNSF).